\documentclass[sigconf]{acmart}

\copyrightyear{2021}
\acmYear{2021}
\setcopyright{acmcopyright}
\acmConference[SBC '21]{Proceedings of the Ninth International Workshop on Security in Blockchain and Cloud Computing}{June 7, 2021}{Virtual Event, Hong Kong}
\acmBooktitle{Proceedings of the Ninth International Workshop on Security in
Blockchain and Cloud Computing (SBC '21), June 7, 2021, Virtual Event, Hong Kong}
\acmPrice{15.00}
\acmDOI{10.1145/3457977.3460301}
\acmISBN{978-1-4503-8405-6/21/06}

\usepackage{enumitem}
\usepackage{diagbox}
\usepackage{multirow}
\usepackage{listings}
\usepackage{bigstrut}
\usepackage{threeparttable}
\usepackage[linesnumbered,ruled,vlined]{algorithm2e}
\usepackage{booktabs}
\usepackage{bm}
\usepackage{arydshln}
\usepackage{framed}

\newcommand{\aave}{\textit{Aave}\xspace}
\newcommand{\dydx}{\textit{dYdX}\xspace}

\newcommand{\uniswap}{\textit{Uniswap}\xspace}

\newcommand{\uniswapTwo}{\textit{UniswapV2}\xspace}

\newcommand{\tab}{\hspace*{1em}}
\newcommand{\code}[1]{{\fontfamily{cmtt}\fontseries{m}\fontshape{n}\selectfont\small{#1}}}
\newcommand{\fl}{Flash Loan}
\newcommand{\defi}{\textit{DeFi}\xspace}

\pdfoutput=1
\begin{document}\sloppy

\fancyhead{}

\title{Towards A First Step to Understand Flash Loan and Its Applications in DeFi Ecosystem}

\author{Dabao Wang$^{1}$,\hspace{0.1em} Siwei Wu$^{1}$,\hspace{0.1em} Ziling Lin$^{1}$,\hspace{0.1em} Lei Wu$^{1}$,\hspace{0.1em} Xingliang Yuan$^{2}$,\hspace{0.1em} Yajin Zhou$^{1}$,\hspace{0.1em} Haoyu Wang$^{3}$,\hspace{0.1em} and Kui Ren$^{1}$}
\affiliation{%
  \institution{$^1$Zhejiang University \& Key Laboratory of Blockchain and Cyberspace Governance of Zhejiang Province; \\ $^2$Monash University; $^3$Beijing University of Posts and Telecommunications; \\
    \{dabao.wang, wusw1020, linziling, lei\_wu\}@zju.edu.cn,
  xingliang.yuan@monash.edu, \\yajin\_zhou@zju.edu.cn, haoyuwang@bupt.edu.cn, kuiren@zju.edu.cn}
}

\begin{abstract}
    {\fl}, as an emerging service in the decentralized finance ecosystem, allows users to request a non-collateral loan. While providing convenience, it also enables attackers to launch malicious operations with a large amount of asset that they do not have. Though there exist spot media reports
    of attacks that leverage Flash Loan, there lacks a comprehensive understanding of existing Flash Loan services. 
    
    In this work, we take the first step to study the {\fl} service provided by three popular platforms. Specifically, we first illustrate the interactions between {\fl} providers and users. Then, we design three patterns to identify {\fl} transactions. Based on the patterns, $76,303$ transactions are determined.
    The evaluation results show that the {\fl} services get more popular over time. 
    At last, we present four {\fl} applications with real-world examples and propose two potential research directions.
\end{abstract}
\keywords{Blockchain; DeFi; Flash Loan}

\maketitle

\section{Introduction}
\label{sec:introduction}

Decentralized finance, aka \textit{DeFi}, has been growing in recent years. Up to 31th Jan 2021, the total value locked (TVL)~\footnote{TVL of a specific protocol represents the total amount of assets staked by users.} in {\defi} has reached 28 billion USD~\cite{defipulse}. 
A service called {\fl} (i.e. uncollateralized loan), which does not exist in the traditional
finance system, has drawn much attention.
However, the introduction of {\fl} is a double-edged sword. 

On the one hand, it does bring in convenience~\cite{defisaver} and facilitate the prosperity of {\defi}. Traders without much capital can launch arbitrage, liquidation and asset swapping with {\fl}. For instance, when traders discover a price difference among tokens between decentralized exchanges (DEXes), they can borrow a considerable amount of capital by {\fl} (no collaterals are required) to maximize the profit.

On the other hand, {\fl} also enables attackers to launch malicious operations with a large amount of capital that they do not have. Therefore, the attack consequences can be vastly amplified. 
In early 2020, two infamous incidents~\cite{qin2020attacking} caused a huge loss to bZx~\cite{bZx}. The hacker took the advantage of {\fl} to manipulate the market price and made considerable profits of $0.83M$ USD and $1.1M$ USD, respectively. Most recently, a hacker borrowed $15M$ DAI~\cite{dai} via {\fl} to gain over $15M$ USD through repeatedly swapping tokens in the EMN~\cite{EMN} pool. 

As such, there is an urgent need to demystify the {\fl} ecosystem and understand the impact of potential security threats. Unfortunately, few studies have been proposed to serve this purpose.
Specifically, previous studies mainly focused on profit optimization~\cite{qin2020attacking,daian2019flash} and oracles~\cite{liu2020first} used by protocols.
To the best of our knowledge, none of them systematically demystified {\fl} and its applications. To provide effective mitigations for {\fl}, we still lack a comprehensive understanding of {\fl}.

In this paper, we take the first step to systematically study {\fl} and further design three patterns to identify {\fl} transactions. Then, we determine all {\fl} transactions until 31st Jan 2021 based on three proposed {\fl} patterns. As a result, $76,303$ {\fl} transactions are identified. To further demystify the behaviors behind {\fl}, we perform an analysis on four {\fl} applications and explain them with real-world examples. 

\smallskip
\noindent \textbf{This paper makes the following contributions:}\tab
\begin{itemize}
    \item We study three {\fl} providers to understand the working process of {\fl}.
    \item We conduct full-chain measurements on {\fl} transactions launched in the {\defi} ecosystem. As far as we know, this is the first work to give a measurement for {\fl} transactions based on real-world data.
    \item We describe four types of {\fl} applications with real-world examples and propose two potential research directions.
\end{itemize}

\noindent \textbf{Paper organization:} The rest of the paper is arranged as follows.
Section~\ref{sec:relatedwork} presents the related work.
Section~\ref{sec:background} elaborates on some concepts in Ethereum and {\defi}.
Section~\ref{sec:flash_loan} illustrates the general idea of {\fl} and explains its working process under different platforms.
Section~\ref{sec:evaluation} conducts an evaluation on {\fl} transactions. 
Section~\ref{sec:behaviors} describes four applications behind {\fl}.
Section~\ref{sec:discussion} proposes two potential research directions.
Finally, we conclude the paper in Section~\ref{sec:conclusion}.

\section{Related Work}
\label{sec:relatedwork}

Kaihua et al.~\cite{qin2020attacking} investigate two existing exploits that happened on 15th and 18th Feb 2020 and present the details of how traders leverage the {\fl} mechanism with the trick of price manipulation to gain profits. They also propose a process to re-boost two exploits via optimized parameters.
Lewis et al.~\cite{gudgeon2020decentralized} leverage {\fl} to execute a governance attack~\cite{governanceAttack} on MakerDAO~\cite{maker}. Moreover, the proposed strategy leads to a theft of 0.5B USD and unlimited mining of DAIs. 
Bowen et al.~\cite{liu2020first} systematically study 4 oracle designs in {\defi} via comparing their price deviations. Besides, they exhibit the potential vulnerabilities existing among 4 oracle designs. 
Kamps et al.~\cite{kamps2018moon} aggregate the information from the existing pump-and-dump schemes among the classic economic and propose a group of patterns with summarised criteria to identify potential pump-and-dump activities in crypto markets. Xu et al. \cite{xu2019anatomy} also investigate 412 pump-and-dump activities to build a model that predicts the pump behavior for all assets exhibiting in DEXes by estimating its pump likelihood. 
Philip et al.~\cite{daian2019flash} present the breadth of arbitrage bots and their profit-making strategies, which optimize users' network latency and pay a high transaction gas fee to win priority gas auctions (PGAs). Furthermore, they highlight that bots' revenue far exceeds the Ethereum block reward and transaction fees. They state that the blockchain consensus stability might be threatened with such high optimization fees. Eskandari et al. ~\cite{eskandari2019sok}  also study the front-running issues across the 25 most active decentralized applications (DApps) on the Ethereum blockchain and summarized their proposed solutions into useful categories.

\section{Background}
\label{sec:background}

The introduction of the blockchain technique~\cite{nakamoto2008bitcoin} has changed the
financial ecosystem in the world. Especially with the invention of Ethereum~\cite{wood2014ethereum}, there has been a wave of developing the decentralized applications (DApps). Smart contracts, as the basis of DApps, enable a transparent environment and become essential components for the development of {\defi}.

\subsection{Common concepts on Ethereum}
To make this work easy to understand, We first introduce a few common concepts in Ethereum.

\smallskip
\noindent{\bf Account.}
Ethereum is an account-centric blockchain system. There are two types of accounts: External Owned Account(EOA) and smart contract account (smart contract in short). The main difference between them is that EOAs are controlled by private keys, and smart contracts are controlled by codes.
Basically, an EOA is created with the generation of the public and private key pair, and a smart contract is always created by an EOA or another smart contract. Both EOAs and smart contracts are identified by their addresses, like \code{0x16431837a35b5469675b2ba5d9b7575d25b721c3}.

\smallskip
\noindent{\bf Digital Currency.}
Ether and the ERC20 token are two main types of digital currencies in Ethereum. Compared to Ether which is supported natively, ERC20 tokens are supported by smart contracts.
Once a smart contract implements the interfaces of ERC20 token standard~\cite{erc20tokenstandard}, then the smart contract can act as an ERC20 token. Moreover, ERC20 tokens can only be transferred by invoking two ERC20 token standard functions: {\it transfer} and {\it transferFrom}.

\smallskip
\noindent{\bf Transaction.}
All actions on the Ethereum blockchain are based on transactions. A transaction have three purposes: transferring Ether, invoking a smart contract function, and deploying a smart contract. According to the circumstance, there exist two types of transactions: {\it external} transaction and {\it internal} transaction. The {\it external} transactions are initiated from EOAs. Alternatively, once a smart contract is invoked within an {\it external} transaction, the {\it internal} transaction will be triggered.
The word "transaction" written individually in the remaining paper indicates the collection of an {\it external} transaction. 

\smallskip
\noindent{\bf Gas Fee.}
Gas is the unit to measure the computational resource used to run operations in Ethereum. To execute transactions on the Ethereum Network requires users to pay a certain amount of Ether (known as gas fee). The gas fee is equal to the value that gas used in the transaction times provided gas price. Since there is no limitation on defining the gas price, users can control the gas fee for their transactions by setting any gas price. Typically, the higher the gas price is, the faster the transaction is verified on the Ethereum blockchain.

\smallskip
\noindent{\bf Function and Event.} 
The smart contract function is identified by the function signature, which is the first four bytes of the hash value (SHA3) of the function name with the parenthesized list of parameter types. If a user sets a function signature in front of a transaction's call data, then the callee smart contract's corresponding function will be invoked. Smart contracts' developers usually leverage the event to record critical information. For example, the ERC20 token standard specifies an event {\it Transfer} to record the spender, receiver and amount of transferred ERC20 tokens. Similarly, an event is identified by the hash value (SHA3) of the event name with the parenthesized list of parameter types. When an event is triggered, a log with an event hash is recorded in Ethereum.

\subsection{Primitives in \defi}
Decentralized finance is a transparent and permissionless finance ecosystem without relying on intermediaries such as banks. In Ethereum, {\defi} is formed with open-source protocols deployed as smart contracts. In the following, we will introduce some primitives in {\defi}.

\smallskip
\noindent {\bf Decentralized Exchange (DEX).}
In the centralized exchanges (CEXes), users entrust their capital to CEXes for trading, and CEXes need to guarantee security. Conversely, trading on DEXes does not require users to provide access to their private keys. Therefore, users can still have full control of their capital.

In particular, there are two main types of DEXes: \textit{Order Book} and \textit{Automated Market Maker} (AMM). 
{\it Order Book} DEXes usually maintain a list of buy and sell orders. They match the pair of orders with a compatible price in their database.
As for AMM DEXes, they maintain various liquidity pools with a designed price calculating mechanism. The price of assets lying in the pool is usually calculated based on its price mechanism and existing liquidity. In comparison, trading in AMM DEXes is more flexible because there is no need for the matching process.

\smallskip
\noindent {\bf Lending.} 
Lending platforms share interests for depositors to lock their capital in the liquidity pool and provide a collateral loan for borrowers. The lending platforms normally require traders to deposit more collateral than the borrowed assets with a certain ratio. Most of the lending platforms design a protection mechanism called liquidation to prevent the potential loss caused by price slippage on traders' deposited collaterals. Once the collateralization ratio ({\it collateral value}/{\it debt value}) is reached, the lending platforms will first sell lenders' collateral with a discount to liquidators. Then, a certain percentage of collateral will be charged to lenders as a penalty. 
We will discuss more details about liquidation in Section~\ref{sec:behaviors}.

\section{flash loan}
\label{sec:flash_loan}

In this section, we first explain the general idea of {\fl}. Second, we elaborate on three famous {\fl} providers and compare their differences in requirements of using. Furthermore, we summarize a {\fl} pattern for each provider to identify {\fl} transactions.

\begin{figure}[t]
    \centering
    \includegraphics[width=1.0\linewidth]{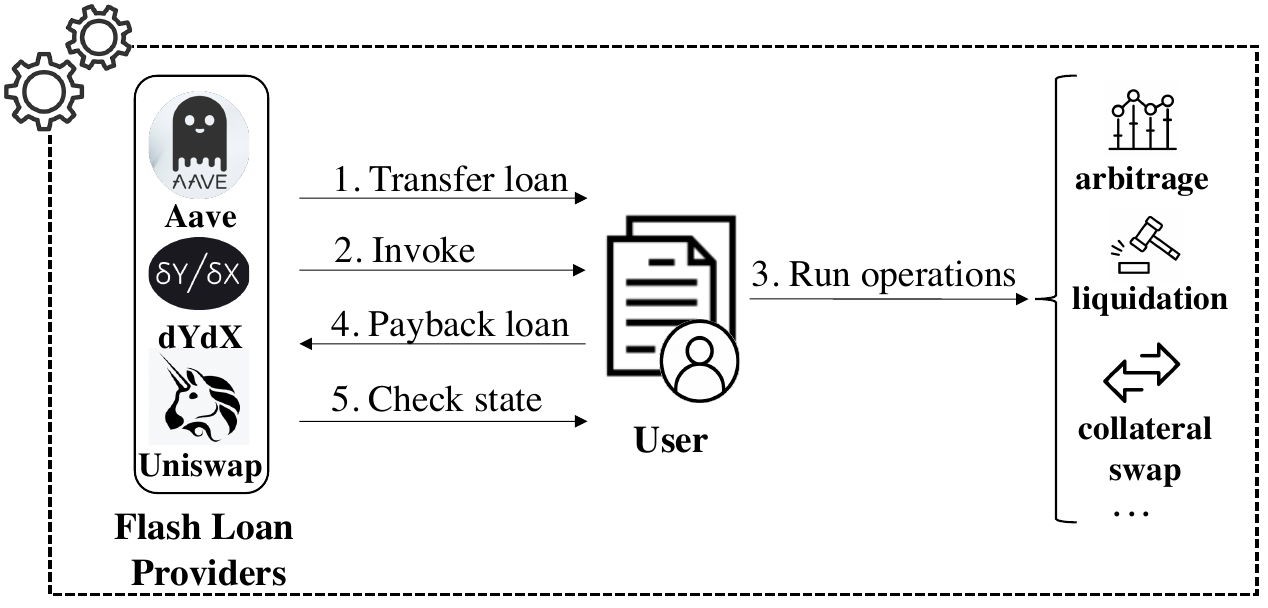} 
    \caption{The workflow of a Flash Loan transaction.} 
    \label{fig:fl_workflow}
    \vspace{-1.5em}
\end{figure}

\subsection{General Idea of Flash Loan}
To request a loan in {\defi} platforms, the user is usually required to deposit overcollateralized assets (i.e., digital cash or tokens). However, a new functionality called {\fl} is developed to enable a non-collateral borrowing service. Moreover, a considerable amount of assets can be ``generously'' lent to users by {\fl} as long as the borrowed assets can be paid back within the current transaction. Otherwise, the platform will instantly revert the transaction to get the lent assets back.

Figure~\ref{fig:fl_workflow} presents the general workflow of a {\fl} transaction. There are two main entities: \textit{Flash Loan Providers} and \textit{Users}. To interact with \textit{Flash Loan Providers}, {\it Users} are required to develop a smart contract. 
A user's contract usually includes three parts: 1) borrowing the loan(s) from {\fl} providers, 2) interacting with other smart contracts, and 3) returning the loan(s). To our best knowledge, there are three main {\fl} providers~\cite{aave}~\cite{dydx}~\cite{uniswap} supporting the {\fl} service with or without certain fees.

Specifically, we generalize the workflow of a {\fl} transaction into five steps. First, \textit{Flash Loan Providers} transfer requested assets to \textit{Users}. Second, they invoke \textit{Users}' pre-designed operations. Third, \textit{Users} will interact with other contracts to execute operations with borrowed assets. Once the execution is completed, \textit{Users} have to return the borrowed assets with or without the extra fee charged by \textit{Flash Loan Providers}. Finally, \textit{Flash Loan Providers} will check their balance. If they discover that no or non-sufficient assets are returned by {\it Users}, they will revert the transaction immediately. Note that all five steps are finished in one transaction.

\subsection{Flash Loan Providers}
\label{subsec:fl_providers}
In this section, we give a basic introduction of each {\fl} provider and reveal their fee-charging mechanisms. Besides, we explain how users' smart contracts interact with different {\fl} providers.

\subsubsection{Aave} 

{\aave}~\cite{aave} is currently the second largest lending platform that locks over $\$3.72B$~\cite{defipulse} up to Jan 2021. As the first platform officially providing the {\fl} service, {\aave} provides a native function called \code{flashLoan}~\footnote{function flashLoan(address \_receiver, address \_reserve, uint256 \_amount, bytes calldata \_params)}
designed in {\aave}'s official contract, i.e., the \code{LendingPool} contract, to trigger {\fl}. Moreover, requesting {\fl} in {\aave} charges $0.25\%$ of the borrowed assets as the fee.

\smallskip
\noindent \textbf{How to prepare flash loan contract with Aave.} For {\aave}'s {\fl}, users need to develop a smart contract consisting of one execution function and one entry-point function. 
The execution function contains users' designed operations for the loaned assets, e.g., trading
in exchanges. Note that, the execution function has to be formed based on 
\code{executeOperation}~\footnote{executeOperation(address \_reserve, uint256 \_amount, uint256 \_fee, bytes calldata \_params)} designed in {\aave}'s official contract \code{FlashLoanReceiverBase}. 

In the entry-point function, users first need to prepare the function \code{flashLoan} to request a loan. Second, users can follow up with the function \code{executeOperation} to run the designed logic on the loaned assets. Third, returning loaned assets must be completed with the provided function \code{transferFundsBackToPoolInternal}~\footnote{transferFundsBackToPoolInternal(address \_reserve, uint256 \_totalDebt)} after finishing executing operations. If {\aave} discover that the vault's state is not balanced (no or non-sufficient assets are paid back to the vault), it will instantly revert the entire transaction.
Once the preparation for the contract is done, users can deploy their contract to the chain and use the {\fl} service from {\aave} by invoking the entry-point function.

\smallskip
\noindent \textbf{Identify flash loan transactions from Aave.}
As aforementioned, {\aave} exposes a native function called \code{flashLoan} for users to utilize {\aave}'s {\fl}. Once the function \code{flashLoan} is invoked successfully, it emits a unique event called \code{FlashLoan}~\footnote{FlashLoan(address indexed \_target, address indexed \_reserve, uint256 \_amount, uint256 \_totalFee, uint256 \_protocolFee, uint256 \_timestamp)}. Therefore, we can use this feature to identify {\fl} transactions from {\aave}. As a result, we discover that there exist over $15,000$ transactions including {\aave}'s {\fl} up to 31st Jan 2021.

\subsubsection{dYdX}
{\dydx}~\cite{dydx} is a non-custodial platform providing services mainly including lending and borrowing on their supporting cryptoassets like ETH, USDC, and DAI. At the time of writing this paper, {\dydx} has locked over $157M$ USD. According to our investigation, there is no native {\fl} feature provided by {\dydx}. However, {\dydx}'s \code{SoloMargin}
contract provides a function called \code{operate}~\footnote{function operate(Account.Info[] memory accounts, Actions.ActionArgs[] memory actions)} that enables to bring a series of operations into one transaction to achieve {\fl} for users. Surprisingly, {\dydx} does not charge any fee for invoking the function \code{operate}.

\smallskip
\noindent \textbf{How to prepare flash loan contract with dYdX.} The preparation for {\fl} in {\dydx} is pretty similar to {\aave}'s. Users are required to develop a contract including one execution function, which contains users' operating logic on the loaned assets, and one entry-point function. In the entry-point function, users first need to sequentially organize a list of provided (by {\dydx}) actions: \code{withdraw}, \code{callFunction}~\footnote{callFunction(address sender, Account.Info memory account, bytes memory data)} and \code{deposit}. Then, users can leverage the function \code{operate} to run the actions one by one to perform {\fl} logic. Note that, \code{callFunction} is acting as the execution function mentioned above to perform users' operating logic. 
In details, \code{withdraw} helps users borrow assets from {\dydx} without any collateral. Then, \code{callFunction} is executed to run users' particular operations on the loaned assets. Finally, \textit{deposit} pays back the loan.
Once the contract is well prepared and deployed on the chain, users can run {\fl} in {\dydx} by invoking the entry-point function.

\smallskip
\noindent \textbf{Identify flash loan transactions from dYdX.}
Though {\dydx} does not directly provide {\fl} feature, users can still achieve {\fl} service in {\dydx} by sequentially executing a series of actions: \textit{Operate}, \textit{Withdraw}, \textit{callFunction}, \textit{Deposit}. Note that, all actions have corresponding event logs: \code{LogOperate}, \code{LogWithdraw}, \code{LogCall}, and \code{LogDeposit}.
Therefore, to identify transactions containing {\dydx}'s {\fl} service, two conditions should be checked. First, all actions' event logs should exist in a transaction. Second, all event logs have to follow a particular order showed below:  
\begin{center}
    \code{LogOperate} $\rightarrow$ \code{LogWithdraw} $\rightarrow$ \code{LogCall} $\rightarrow$ \code{LogDeposit}
\end{center}

Once two conditions are both satisfied in a transaction, we can confirm that it is a {\fl} transaction from {\dydx}. Based on our experiment, around $25,000$ transactions are identified as {\fl} transactions leveraging {\dydx}'s service.

\subsubsection{UniswapV2}
As one of the most famous DEX protocols, \textit{Uniswap}~\cite{uniswap} occupies around $11\%$ ($3.2B$ USD) of liquidity in {\defi} ecosystem. Different from {\aave} and {\dydx}, \textit{Uniswap} simply builds its {\fl} feature called \textit{flash swap} on the function \code{swap}~\footnote{swap(uint amount0Out, uint amount1Out, address to, bytes calldata data)}. In details, the function \code{swap} switches in between flash swapping and normal swapping based on the provided parameters. In terms of the fee, compared to the aforementioned {\fl} providers, {\uniswapTwo} charges the highest fee ($0.3\%$) based on users' borrowed assets. 

\smallskip
\noindent \textbf{How to prepare flash loan contract with UniswapV2.} A contract for using {\fl} in {\uniswapTwo} requires users code their designed operations in function \code{IUniswapV2Callee} which inherits from {\uniswapTwo}'s interface standard 
\code{IUniswapV2Callee}
. 
The designed operations must include repayment action to success {\it flash swap}. Unlike the two {\fl} providers previously analyzed, users do not need to develop any entry-point function to initiate a transaction. Instead, users first neet to find the targeted pair contract~\footnote{Uniswap has many pair contracts, which are published for users, to launch a swap on a pair of tokens. In {\uniswapTwo}, every pair contract supplies \code{swap} function.} published by {\uniswap}. 
Then, through invoking the function \code{swap} with specific parameters, a {\fl} transaction will be triggered. In particular, parameter \code{to} should be the address of the deployed contract, and the length of parameter \code{data} should be greater than zero.

\smallskip
\noindent \textbf{Identify flash loan transactions from UniswapV2.}
Identifying {\fl} transactions from {\uniswapTwo} requires three steps.
First, we verify the event \code{PairCreated} emitted by the \textit{UniswapV2Factory}
contract and collect a group of pair contracts (addresses) that supplies the \code{swap} function. 
Second, we verify the event \code{swap} emitted by triggering the function \code{swap} in all transactions.
Lastly, once we confirm that the transaction invokes the \code{swap} function of pair contracts, we identify it as {\fl} transaction from {\uniswapTwo} through checking three conditions: 
\begin{enumerate}
    \item The length of the parameter \textit{data} is greater than zero.
    \item The internal transaction triggered by \code{uniswapV2Call} must include the invocation of \code{transfer} or \code{transferFrom} function.
    \item The receiver address of \code{transfer} or \code{transferFrom} function must be the pair contract.
\end{enumerate}
In conclusion, if all three conditions are fulfilled, the transaction can be confirmed as a operation of \textit{flash swap} (i.e., {\fl} in {\uniswapTwo}). Through applying the identifying pattern, around $36,500$ transactions are filtered.

\smallskip
Importantly, since our patterns strongly rely on the specific rule published by each {\fl} provider, the identified {\fl} transactions will not result in any false positive.

\section{The Measurement of Flash Loan Transactions}
\label{sec:evaluation}
In this section, we perform a measurement for {\fl} transactions. 
As for the data set, we collect about one billion transactions (up to 31th Jan 2021) from the Ethereum blockchain ledger. 

\begin{table}[t]
	\caption{The distribution of {\fl} transactions.}
	\label{tbl:dis_fl}
	\centering
	\begin{tabular}{l|l|l|l}
		\hline        		
		{\bf Providers} & {\bf \# Transactions}   & {\bf \# Receivers} & {\bf \# Average Use}   \\
		\hline
		\hline
		{\it Aave}      & $15,016$   & $463$ & $32.4$ \\
		\hline
		{\it dYdX}      & $24,983$  &  $666$ & $37.5$ \\
		\hline
		{\it UniswapV2} & $36,574$   & $346$ & $105.7$ \\
		\hline
 	\end{tabular}

 \vspace{-10pt}
\end{table}

\smallskip
\noindent \textbf{Statistic Result.} Through applying three {\fl} patterns proposed in Section~\ref{sec:flash_loan} over one billion collected transactions, $76,303$ {\fl} transactions are identified.
As shown in Table~\ref{tbl:dis_fl}, users leverage {\fl} in {\uniswapTwo} most frequently, as there are $36,574$ ($48\%$) transactions. Following up, $24,983$ ($32.5\%$) {\fl} transactions are found in {\dydx}, and $15,016$ ($19.5\%$) {\fl} transactions are found in {\aave}. The second last column of Table~\ref{tbl:dis_fl} records the amount of {\fl} receivers for each {\fl} provider. According to the results, {\fl} in {\dydx} has been used by most unique receivers ($666$) as well as there are $463$ and $346$ {\fl} receivers in {\aave} and {\uniswapTwo}, respectively. In the last column of Table~\ref{tbl:dis_fl}, it presents the average number of transactions launched by each receiver.

\begin{figure}[t]
    \centering
    \includegraphics[width=1.0\linewidth]{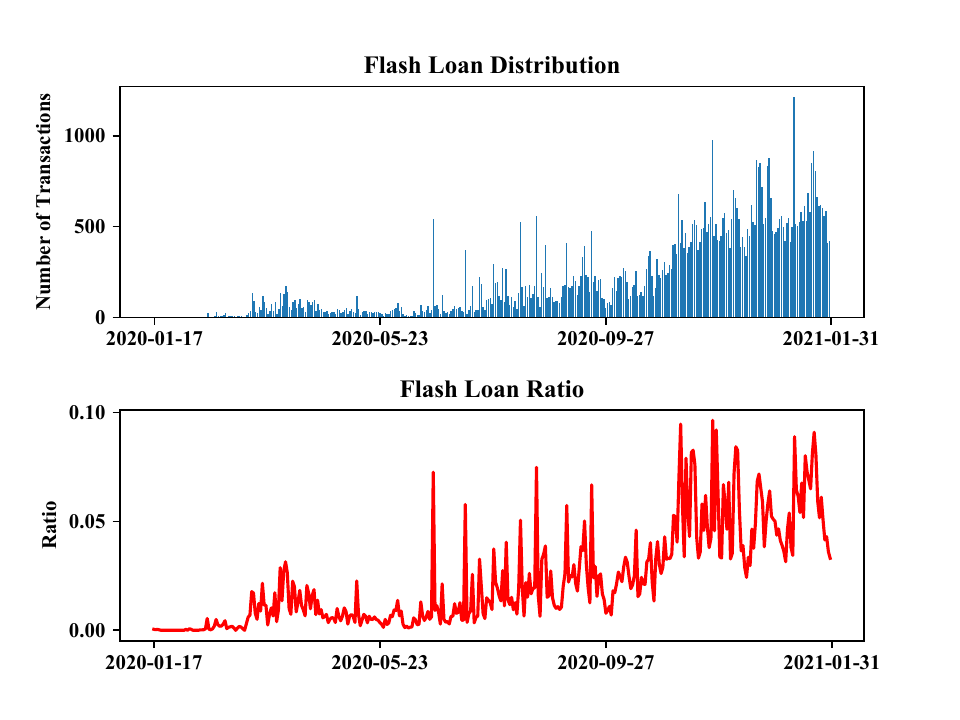} 
    
    \caption{The distribution of Flash Loan transactions.} 
    \label{fig:fl_dist} 
    \vspace{-1.5em}
\end{figure}

\smallskip
\noindent \textbf{Trend.} To further understand the popularity of {\fl} in {\defi}, we provide two time series for the number and the ratio (the number of {\fl} transactions / the number of all transactions) of {\fl} transactions in Figure~\ref{fig:fl_dist}. Note that, both the number and the ratio of {\fl} transactions are increasing over the time.

\smallskip
\noindent \textbf{Finding.} Nowadays, {\fl} service is getting more popular. Through measuring {\fl} transactions, we discover that {\fl} in {\uniswapTwo} is used most intensively ($105.7$ transactions per receiver), while {\fl} in {\aave} ($32.4$ transactions per receiver) is used least intensively.

\section{Applications Of Flash Loan}
\label{sec:behaviors}
{\fl} can be used for legitimate purposes such as arbitrage, liquidation, etc. Besides, {\fl} can also become a sharp knife for aggressive users to harm the {\defi} ecosystem. In this section, we describe four applications of {\fl} and discuss the benefit that {\fl} brings to them.

\subsection{Arbitrage}
General speaking, arbitrage in {\defi} is a behavior to gain benefits by trading in between platforms supplying different price for an asset. Since the {\defi} market reacts slower for events happening in the network than the real-world market, traders can take advantage of the market's inefficiencies to buy and sell the cryptoassets at a different price to gain financial benefits. Note that, arbitrage itself
is not a malicious behavior. In fact, the arbitrage can be leveraged to balance the token prices between DEXes. 

\smallskip
\noindent \textbf{Benefit.} With {\fl}, traders can launch arbitrage without any pre-owned asset. In particular, if the price difference is found, the arbitrageurs can instantly borrow a considerable asset with {\fl} service to earn benefits. Therefore, arbitrages with {\fl} become ``cost-free'' as long as traders can afford the gas fee to launch the transaction.

\smallskip
\noindent \textbf{Example.} We will present an arbitrage that happened on 22nd Jan 2021 with four steps. First, in the transaction~\footnote{0x2a0c2599f89d95a46f4f28712e99a71847d8f72af5bdc8942d6c9dd01d896624}, the trader borrowed 1.13 Ether from {\dydx}. Second, 1.13 Ether was converted to 345 LPT tokens in \textit{Balancer}~\cite{balancer} through a trade. Third, another trade was triggered to trade 345 LPT tokens on the corresponding liquidity pool of {\uniswap}. As a result, 1.46 Ether were gained by the trader. Finally, the trader returned 1.13 Ether to {\dydx} from gaining. In this arbitrage case, the trader gained 0.33 Ether (around 538 USD at the time) by paying 0.05 Ether as the gas fee.

\subsection{Wash Trading}
Wash trading in {\defi} is a behavior that creates fake trading volume for certain cryptoassets or platforms. Specifically, wash trading is a group of trades increasing the trading volume on the asset or platforms. In reality, wash trading can easily mislead users to perform financial operations on the targeted cryptoassets and platforms. Though some countries like the U.S. have banned washing trading to protect their traditional markets and the stock market, it is brought back to the crypto market again because of the popularity of cryptocurrency and the lack of legal management. 

\smallskip
\noindent \textbf{Benefit.} With {\fl}, wash traders can manipulate the market without a large amount of capital as long as they can afford the potential loss and the gas fee.

\smallskip
\noindent \textbf{Example.} In the transaction~\footnote{0x8fc77fa516aca91715046c1f307397ac49d211244fced5734c480a660015f927} executed on 14th Jul 2020, the trader firstly borrowed 10 Ether from {\aave}. Then, five repeated trades were launched to increase the trading volume in the liquidity pool {\it ``Uniswap V2: DAI 2''}. In details, for each trade, 2 Ether was first converted to \textit{DAI} and all {\it DAI} would be instantly converted back to Ether at the pool. 
Finally, the user paid back the {\fl}. There were no further operations except wash trading in this transaction. As a result, the trader lost 0.068 Ether in this transaction and paid 0.164 Ether as the gas fee. 

\subsection{Flash Liquidation}
Liquidation is a behavior launched by the liquidator to buy undercollateralized assets from the lending platforms.
There are two liquidation classes ({\it Fixed Price Biding} and {\it Auction}) involving three roles (platforms, liquidators and collateral keepers). 
For fixed price biding, the lending platforms like {\dydx} and \textit{Compound} allow liquidators to buy undercollateralized assets from collateral keepers with a specific discount. Moreover, the lending platforms will apply a fixed amount of liquidation penalty to collateral keepers. Alternatively, the platforms like MakerDAO~\cite{maker} allow liquidators to compete on the keeper's undercollateralized assets like an auction. The winners, who pay the higher gas fee to launch their transactions, can buy the undercollateralized collateral with a discount.

\smallskip
\noindent \textbf{Benefit.} With {\fl}, anyone can become a liquidator to make profits without much capital by buying the undercollateralized assets with a specific discount. 

\smallskip
\noindent \textbf{Example.} The transaction~\footnote{0x38b706beda9426027081f2b6c1f2d6e68b2387d824a59e20a4d6decdfec43385} performed a liquidation on 3rd Nov 2020. In details, the liquidator first borrowed $12,940$ \textit{DAI} from {\dydx} and swapped the {\it DAI} to $13,046$ \textit{USDT}~\cite{tether}. Second, $13,046$ \textit{USDT} was used to buy the asset from the undercollateralized position in \textit{Compound}~\cite{Compound}. Through exchanging the bought asset, the liquidator got $13,450$ \textit{DAI}. Finally, after paying back the {\fl}, $510$ \textit{DAI} (about $510$ USD) remained as profits, which is greater than the gas fee (about $172$ USD).

\subsection{Collateral Swap} 
Collateral swap in {\defi} is a well-defined behavior consisting of two main steps: 
\begin{itemize}
    \item \textit{Swapping:} Redeeming the collateral from the old loan.
    \item \textit{Operating:} Lauching operations on redeemed collateral. 
\end{itemize}
Since the crypto market is extremely unpredicted, timely closing existed collateral position becomes an urgent need for the holder to stop loss from severe slippages and liquidations.

\smallskip
\noindent \textbf{Benefit.} For users without sufficient capitals for \textit{Swapping}, {\fl} can solve their urgent need by providing ``cost-free'' assets to save their collaterals from the price slippage and the liquidation. Besides, {\fl} also enables \textit{Swapping} and \textit{Operating} actions run within one transactions. It further prevents users from suffering the uncertainty (like slippage) between transactions. 

\smallskip
\noindent \textbf{Example.} In the transaction~\footnote{0xaf4ca18a0d3b94d948a9eeb47ba57c84c212aaeb7284b38ede6a0f6f549c3827} launched on 17th Mar 2020, the user performed a collateral swap to take out \textit{BAT} (\textit{Swapping}) and exchanged it to more stable asset \textit{USDC} (\textit{Operating}). As the context, the user previously opened a loan of \textit{DAI} by depositing the asset \textit{BAT}. In details, first, the user borrowed 25 \textit{DAI} {\fl} from {\aave}. Second, 25 \textit{DAI} was paid to redeem out $504$ \textit{BAT} (collateral). Third, the redeemed collateral was further swapped to a more stable asset \textit{USDC}. Lastly, the user converted part of the \textit{USDC} to pay back {\aave}'s {\fl}. As a result, the user swapped his/her collateral to a more stable asset without holding any \textit{DAI}.

\smallskip
In conclusion, {\fl} provides a vast convenience for multiple applications (arbitrage, wash trading, liquidation, and collateral swap) in {\defi} ecosystem. It can either bring traders benefits or be used maliciously. Thus, understanding the intention of the application is necessary for both traders and developers.

\section{Future Research Directions}
\label{sec:discussion}

In this section, we propose two potential research directions.

\smallskip
\noindent \textbf{Arbitrage.}
Arbitrages in {\defi} happen nearly every day. By leveraging the smart contract and {\fl}, many organizations and individuals create bots to launch designed operations. We believe that the arbitrage bots in {\defi} can maximize traders' profits if the information (i.e. price difference) can be timely detected and fed to the bot.

\smallskip
\noindent \textbf{DeFi Attacks.}
With the increasing popularity of {\defi}, attackers could steal money from {\defi} platforms or individuals. Identifying malicious transactions, especially the zero-day attacks, is challenging due to complicated interactions between multiple entities (Figure~\ref{fig:fl_workflow}). How to propose effective methodologies to detect attacks towards DeFi platform is still an open research question.
 
\section{Conclusion}
\label{sec:conclusion}

This paper takes the first step to study the working process of {\fl} within three different platforms. In this work, we identified $76,303$ {\fl} transactions and $1,454$ {\fl} receivers. Furthermore, we evaluated the popularity of {\fl}. To better understand the application behind the {\fl} mechanism, we elaborated on four types of applications with real-world examples. Finally, we proposed two potential research directions in this area.

\section*{Acknowledgments}
We thank the anonymous reviewers for the thorough reviews. This work was supported by the Leading Innovative and Entrepreneur Team Introduction Program of Zhejiang (2018R01005), and the Fundamental Research Funds for the Central Universities (K20210226).

\bibliographystyle{ACM-Reference-Format}
\bibliography{sbc2021}
\end{document}